\documentclass[12pt,preprint]{aastex}
\doublespace

\usepackage{epsfig}
\usepackage{natbib}
\usepackage{wasysym}
\usepackage{subfigure}
\usepackage{verbatim}
\usepackage{enumerate}
\usepackage{graphicx}

\usepackage{color}

\slugcomment{ }

\shorttitle{Rotation in Tornado-like Prominence}
\shortauthors{Su et al.}

\begin{document}

\title{Solar Magnetized Tornadoes: Rotational Motion in a Tornado-like Prominence}

\author{Yang Su\altaffilmark{1}}
\email{yang.su@uni-graz.at}
\author{Peter G\"{o}m\"{o}ry\altaffilmark{2}}
\author{Astrid Veronig\altaffilmark{1}}
\author{Manuela Temmer\altaffilmark{1}}
\author{Tongjiang Wang\altaffilmark{3,4}}
\author{Kamalam Vanninathan\altaffilmark{1}}
\author{Weiqun Gan\altaffilmark{5}}
\and
\author{YouPing Li\altaffilmark{5}}


\altaffiltext{1}{IGAM-Kanzelh\"{o}he Observatory, Institute of Physics, University of Graz, Universit\"{a}tsplatz 5, 8010 Graz, Austria.}
\altaffiltext{2}{Astronomical Institute of the Slovak Academy of Sciences, SK-05960 Tatransk\'{a} Lomnica, Slovakia}
\altaffiltext{3}{Department of Physics, the Catholic University of America, Washington, DC 20064, USA.}
\altaffiltext{4}{Solar Physics Laboratory (Code 671), Heliophysics Science Division, NASA Goddard Space Flight Center, Greenbelt, MD 20771, USA.}
\altaffiltext{5}{Key Laboratory of Dark Matter and Space Astronomy, Purple Mountain Observatory, Chinese Academy of Sciences, Nanjing 210008, China.}

\begin{abstract}
\cite{2012ApJ...756L..41S} proposed a new explanation for filament formation and eruption, where filament barbs are rotating magnetic structures driven by underlying vortices on the surface. Such structures have been noticed as tornado-like prominences when they appear above the limb. They may play a key role as the source of plasma and twist in filaments. However, no observations have successfully distinguished rotational motion of the magnetic structures in tornado-like prominences from other motions such as oscillation and counter-streaming plasma flows. Here we report evidence of rotational motions in a tornado-like prominence. The spectroscopic observations in two coronal lines were obtained from a specifically designed Hinode/EIS observing program. The data revealed the existence of both cold and million-degree-hot plasma in the prominence leg, supporting the so-called ``the prominence-corona transition region''. The opposite velocities at the two sides of the prominence and their persistent time evolution, together with the periodic motions evident in SDO/AIA dark structures, indicate a rotational motion of both cold and hot plasma with a speed of $\sim$5 km s$^{-1}$. 
\end{abstract}
 

\section{Introduction}

Filaments are cold and dense plasma clouds suspended in the hot corona, stabilized by their magnetic field. They are called prominences when they appear above the solar limb. Tornado-like prominences \citep{1943ApJ....98....6P,1950PASP...62..144P} are one type of prominences that show tornado-like vertical structures. However, whether they are as same as the recently discovered solar tornadoes \citep{2012Natur.486..505W} is still a question \citep{2013ApJ...774..123W}. High-cadence EUV imagery of SDO/AIA revealed new details of tornado-like prominences \citep{2012ApJ...752L..22L,2012ApJ...756L..41S,2013A&A...549A.105P}. The phenomenon regained much attention, since it may be intimately linked to the formation of filaments as well as to instabilities leading to filament eruption \citep[e.g.][]{2012ApJ...756L..41S,2013ApJ...774..123W}.
 

\cite{2012ApJ...756L..41S} studied a tornado-like prominence in H-alpha, SDO and STEREO images, finding that those vertical structures are actually barbs of filaments and they appear to be rotating. They pointed out the possible connections among three phenomena, vortices \citep[e.g.][]{1988Natur.335..238B,2008ApJ...687L.131B,2009A&A...493L..13A}, filaments and solar tornadoes. The role of vortices in generating rotating coronal structures has been discussed in \cite{1999SSRv...87..339V} and \cite{2012Natur.486..505W}. By tracking the magnetic elements in the photosphere, \cite{2012ApJ...756L..41S} identified a vortex under one of the tornado-like barbs. However, the method they used \citep[DAVE,][]{2006ApJ...646.1358S} works best for strong magnetic fields. Therefore, the weak field strength (around 20 to 30 Gauss) in the studied region may not result in solid evidence. The paper lacked convincing evidence that these coronal structures in filament barbs are indeed rotating, which is important for the vortex-filament picture they proposed. Although rotating magnetic structures on solar surface have been observed before \citep[the ``solar cyclones'',][]{2011ApJ...741L...7Z}, the connection between them and filaments are not clear. 

Meanwhile, other independent works provided supports for this interpretation. \cite{2012Natur.486..505W} detected individual solar magnetic tornadoes in the solar atmosphere. Their observations and simulation provided evidence that solar tornadoes are indeed driven by the underlying vortices. However, the structures they detected have smaller scale and much shorter lifetime, compared with tornado-like prominences. In a follow-up study, \cite{2013ApJ...774..123W} mapped the velocity field of a tornado-like prominence in H-alpha data and found a red-shifted and a blue-shifted region on the two sides of the prominence leg. This was a further step to identify the rotational motion previously detected along a single slit in prominence legs at He {\sc I} 10830 {\AA} \citep{2012ApJ...761L..25O}. 
  
These results suggest that plasma at the two sides of tornado-like prominence or filament barbs move oppositely in the line of sight (LOS), which is expected for a rotating structure. Actually, such red/blue shifts are also found at the two sides of other structures, such as spicules, and generally taken as evidence of rotational motion \citep{1998SoPh..182..333P, 2011AA...532L...9C}. However, a snap shot of the plasma Doppler velocity is still not sufficient to distinguish a rotational motion from other motions. According to the interpretations of \cite{2012ApJ...756L..41S} and \cite{2013ApJ...774..123W}, the magnetic structures in solar tornadoes rotate as a whole and the plasma within can move both along magnetic field lines and together with the rotating magnetic field. \cite{2013SoPh..tmp..189P} argued that the apparent rotation of tornado-like prominences is merely caused by oscillation, plasma flow along helical structure or projection effects. Indeed, oscillation in filament/prominences have been widely reported and well studied \citep[e.g. reviews by][]{2011SSRv..158..237L,lrsp-2012-2}. It is one of the main processes that could cause the ``illusion'' of rotation. Counter-streaming flows in filaments \citep{1998Natur.396..440Z,2013ApJ...775L..32A} may also result in red/blue shifts in Doppler maps. 

%


The key question here is whether the magnetic structures in tornado-like prominences are indeed rotating. However, the identification of rotational motions of magnetic structures cannot be done in pure imaging data but needs spectroscopic Doppler observations, their evolution, and, ultimately, direct measurements of coronal magnetic field. Oscillation and rotation may cause a similar red/blue pattern in velocity map but a different time evolution. Oscillation would result in alternate changes in the signs of Doppler velocities on both sides of the rotation axis, whereas rotation leads to persistent velocity pattern. Therefore, continuous Doppler measurements with time spanning more than one rotation/oscillation period are essential for the identification. In order to obtain both spatial structure and temporal evolution of the plasma velocity, we proposed a Hinode/EIS observing program which was specially designed to study plasma motions and evolution in tornado-like prominences. The program consists of a sequence of two different observing modes, a scanning mode and a sit-and-stare mode. The former one runs before and after the latter one, which lasts for about three hours. The program was assigned as HOP 0237 \footnote[1]{For more details see: \url{http://www.isas.jaxa.jp/home/solar/hinode$\_$op/hop.php?hop=0237}} and performed for the first run during September 9\,--\,16, 2013. In this paper, we analyze a tornado-like prominence observed at the west limb on 14 September 2013 between 03:02 and 07:28\,UT.

%
 

\section{Data and data reduction}

The EUV Imaging Spectrometer \citep[EIS,][]{2007SoPh..243...19C} on-board the Hinode satellite provides simultaneous spectroscopic observations in the wavelength ranges 170-210 {\AA} and 250-290 {\AA}. The spectral lines located in these two bands cover a temperature interval log\,T\,=4.7-7.3. EIS's spectral resolution of 0.0223 {\AA} pixel$^{-1}$ allows us to measure Doppler shifts with a precision in the order of a few km s$^{-1}$. The pixel size is about 1${\arcsec}$.

In the EIS scanning mode, a raster image of the 2D region with a FoV of 100${\arcsec}$\,$\times$\,256${\arcsec}$ is taken to map the selected part of the solar atmosphere. The 2${\arcsec}$ wide slit is used to scan the selected region in 49 steps (i.e. 50 exposures are taken). In the sit-and-stare mode, the spectrograph slit (width: 2${\arcsec}$, length 256${\arcsec}$) is placed to the center of the previously taken raster to measure time evolution of the spectroscopic parameters in a small sub field of the scanned region. A constant exposure time of 50\,s is used for both observing modes.

All EIS data were calibrated with the EIS$\_$PREP routine in SolarSoftWare (SSW) using the standard processing options. The data were also corrected for spatial offset which exists between signals detected at short and long wave bands. The observed spectral profiles were approximated with Gaussian functions in order to determine the spectral parameters, i.e. integrated intensities, Doppler shifts and line widths. The zero reference of the Doppler shifts were calculated from the average value of the particular Doppler shifts from quiet Sun regions.

EIS jitters, which may cause shifts in both X and Y directions of the EIS slit, could not be corrected (the alignment file is not available at this point). We minimized possible shifts by cross-correlating between intensities along the EIS slit and a sub field extracted from the AIA 193 {\AA} image. Both data are taken at ~03:50 UT (start of the sit-and-stare mode). Although the actual location of the EIS slit may change in time by a few arcseconds, this does not change our main findings and conclusions.

The prominence was best observed in Fe\,{\sc xii}\,195.12 {\AA} and Fe\,{\sc xiii}\,202.04 {\AA} (formation temperature 1.3 and 1.6 MK, respectively) among our selected lines. The Fe\,{\sc xii}\,195.12 {\AA} is blended by other Fe\,{\sc xii} emission at 195.18\,\AA. Therefore we used a double-Gaussian function to fit the measured profiles of each pixel. In order to get more robust results we tied the spectral parameters of both components in the following way: the spectral widths of both components had to be equal (both components are emitted by the same ion) and the positions of the centroids of both components were always separated by 0.06 {\AA} (difference of the laboratory wavelengths of both components). The Fe\,{\sc xiii}\,202.04 {\AA} is not blended. Therefore a single Gaussian function was applied here.

The target was also observed in EUV by the Atmospheric Imaging Assembly \citep[AIA, ][]{2012SoPh..275...17L} on board the Solar Dynamics Observatory (SDO). AIA observes the full Sun in ten UV and EUV passbands covering temperatures from $\sim$5,000 K to $\sim$20 MK with high temporal cadence ($\sim$12 sec) and pixel size ($\sim$0.6$\arcsec$). Here a selected image-set with 24 sec cadence was used to better understand the spatial structure and evolution of the prominence. 

 
\section{Results}

Fig. 1 presents snapshots of the tornado-like prominence under study in different wavelengths: EIS Fe {\sc xii} 195 {\AA} (scan mode), AIA 171 {\AA}, 193 {\AA}, AIA 304 {\AA}, and H-alpha line center from the Solar Magnetic Activity Research Telescope (SMART) at Hida Observatory, Kyoto University. The vertical, dark structures visible in EUV above the limb are the barbs of the filament, whose spine is connecting the top of barbs and the structures in the south (see Fig. 1 and the online movie). The spine has a height of at least 20$\arcsec$ (~15 Mm), and is therefore much larger than the simulation scale ($\sim$3 arcsec) of solar tornadoes in \cite{2012Natur.486..505W}. As other prominences, this prominence emits in cool lines, such as AIA 304 {\AA} and H-alpha, but absorbs/blocks the background emission in coronal lines like EIS Fe {\sc xii} 195 {\AA} and AIA 193 {\AA}, resulting in the appearance of dark structures in EUV \citep{2010SSRv..151..243L}. 

The EIS Fe {\sc xii} 195 {\AA} intensity and LOS velocity map of the whole region derived from the EIS scan mode are shown in the top row of Fig. 2. Detailed distributions of the intensities and velocities along two selected lines across the vertical structure are plotted in the bottom row. These parameters are obtained from the spectral fitting to the line profiles of each pixel (see Fig. 3 for more details of the fitting results). The velocity map shows clearly a red-shifted region in the northern side and blue-shifted region in the southern side of the tornado-like prominence. A smooth change from red to blue is found in between. No clear structure is seen in the map of non-thermal velocities (top right plot in Fig. 2) above the background level, suggesting a minor effect on the line broadening from multiple thin threads \citep{2005SoPh..226..239L} along LOS.   

Fig. 3 shows the actual line profiles of EIS Fe {\sc xii} 195 {\AA} and the spectral fitting results for six selected pixels (P1 to P6), three in the red-shifted zone and three in the blue-shifted zone. Particularly, the fitting functions for P1 and P4 are directly compared in the bottom left plot. It clearly shows the red-shifted and blue-shifted line profiles. 

Fig. 4b shows the time-distance map for the intensities along part of the EIS slit obtained in sit-and-stare mode from 03:45 UT to 06:45 UT. The location of the slit is indicated in the images taken in the scan mode (Fig. 4a,c,e). Fig. 4d,f are the same maps but for the velocities obtained at EIS Fe {\sc xii} 195 {\AA} and Fe {\sc xiii} 202 {\AA}. A movie of AIA 193 {\AA} images and EIS velocities along the slit is available online. The time-distance maps of velocity clearly show the persistent red and blue shifts, indicating that the opposite signs of velocities on each side of the dark structure do not change with time. The black lines in the time-distance maps connect the darkest pixels for each exposure interval. They are found to be located between red-shifted plasma and blue-shifted plasma, where the velocities are close to zero. These results give strong evidence that the red/blue shifts are not due to oscillation along the LOS but due to rotational motion of the plasma.

Fig. 4h is the time-distance map of AIA 193 {\AA} intensity, obtained at a close location to the 2-arcsec-wide EIS slit (3$\arcsec$ to the west of the EIS slit). This map best shows the periodic motions (the two green dashed lines), which suggest that the cold, dark structures are either rotating or swaying in the plane of sky. The period ($T$) is $\sim$66 minutes and amplitude ($r$) $\approx5\arcsec$. Considering the angle between the slit and the vertical structure, we here use 4.2$\arcsec$ as the actual amplitude ($r_*$). Assuming that they are rotating threads in barbs, we can estimate the rotational linear speed $v=2{\pi}r_*/T\approx$ 5 km s$^{-1}$. This independent measurement is in good agreement with the varying velocities of 2-5 km s$^{-1}$ at the two sides of the structure derived from EIS spectra (Fig. 4c{\sbond}f), giving strong support to the existence of rotating thread-like structures in the prominences. In ideal case, we expect to see clear thread-like velocity patterns following the motion of each thread in the time-distance maps. In fact, as shown in Figure 4d,f, such patterns are indeed observed (i.e., the velocities reverse signs from one side to the other along the two green dashed lines), but they lack details possibly due to the lower spatial resolution of EIS, compared to AIA.

\section{Summary and Discussion}

With our specially designed observing plan, we obtained not only the velocity map that shows spatially resolved red-shifted and blue-shifted regions, but also the time evolution of plasma velocity in the two regions. The persistent red/blue patterns at the two sides of the tornado-like prominence suggest that the plasma therein is rotating around its center axis. Moreover, the Doppler velocities are in agreement with that estimated from the periodic motions seen in AIA image sequence. All these results, spatial distribution of plasma velocity, time evolution of plasma velocity, and the evolution of spatially resolved structures, support that the structures are rotating together with the plasma within. Neither oscillation nor counter-streaming flows alone can explain all these aspects. If the magnetic pressure dominates over the plasma pressure (low plasma beta) in the tornado structure, then the results also indicate that the magnetic structures rotate with the emitting plasma, as shown by the simulation in \citep{2012Natur.486..505W}.

The detected rotational motions in barbs are not part of general models for filament formation but give a strong support to the vortex-filament picture. In this new idea, vortex flows on the surface first converge plasma and magnetic field. The motions also twist the magnetic field lines connecting these vortices to form a group of magnetized tornadoes (Fig. 5a,b). Plasma in the photosphere may be transported upwards into the helical magnetic structures. The vertical parts then become visible on disk as filament barbs and the horizontal part as filament spine, a highly twisted flux rope (Fig. 5c). With greater and greater helicity being built up, the system may be unstable and erupt eventually \citep[Fig. 5d, see observations in][]{2012ApJ...756L..41S,2013ApJ...774..123W}. This is a different approach to form a filament from other models, such as ``coronal condensation'' \citep[see][and references therein]{2012ApJ...745L..21L} and flux rope formation via either magnetic reconnection or flux emergence \citep[see the review in][]{2010SSRv..151..333M}. But it is possible that they apply to different types of filaments.

We also found evidence of million-degree plasma in the prominence. The existence of a hot prominence-corona transition region has been proposed long ago \citep[e.g.][]{2001SoPh..202..293E,2010SSRv..151..243L} but few evidence has been obtained due to the difficulty in separating the foreground coronal emission. The observations presented here were obtained in emitting hot coronal lines, a major difference from previous works which were done at much lower temperatures, such as the He {\sc I} 10830 {\AA} \citep{2012ApJ...761L..25O} and H-alpha spectral lines \citep{2013ApJ...774..123W}. The H-alpha observations in \cite{2013ApJ...774..123W} show that tornado structure may have temperatures in the range from 5000 to 15,000 K. But plasma at this temperature would cause continuum absorption \citep{1998SoPh..183..107K,2005ApJ...622..714A,2010SSRv..151..243L} in the background corona emission, which does not alter the properties (Doppler shifts and shapes) of coronal lines. Therefore, the detected red-shifted and blue-shifted coronal lines mainly formed in the moving hot plasma. Their velocities are also in agreement with that deduced from cold plasma - the motion of dark structures evident at AIA 193 {\AA}, indicating that hot and cool plasma co-exist in the rotating structures. 


\acknowledgments

The H-alpha data was taken by the Solar Magnetic Activity Research Telescope (SMART) at Hida Observatory, Kyoto University.
We downloaded the data from the SMART T1 Data Archive (http://www.hida.kyoto-u.ac.jp/SMART/T1.html). SDO is a mission for NASA's Living With a Star (LWS) Program. 
This work was supported by the project of
the \"{O}sterreichischer Austauschdienst (OeAD) and the
Slovak Research and Development Agency (SRDA)
under the grant numbers SK 16/2013 and SK-AT-0003-12.
Y.S., A.V., M.T., and K.V. acknowledge the Austrian Science Fund (FWF): P24092-N16 and V195-N16. 
Y.S. and A.V. also acknowledge the European Community Framework Programme 7, High Energy Solar Physics Data in Europe (HESPE), grant agreement no.: 263086. 
P.G. acknowledges the support from grant VEGA 2/0108/12 of the Science Grant Agency and from the project of the Slovak Research and Development Agency under the contract No. APVV-0816-11. 
The work of T.W. was supported by NASA grant NNX12AB34G and the NASA Cooperative Agreement NNG11PL10A to CUA.
Y.L. and W.G. acknowledge 11233008 from NNSFC and 2011CB811402 from MSTC.



\clearpage

 \begin{figure}
 \begin{center}
\includegraphics[angle=90,scale=.60]{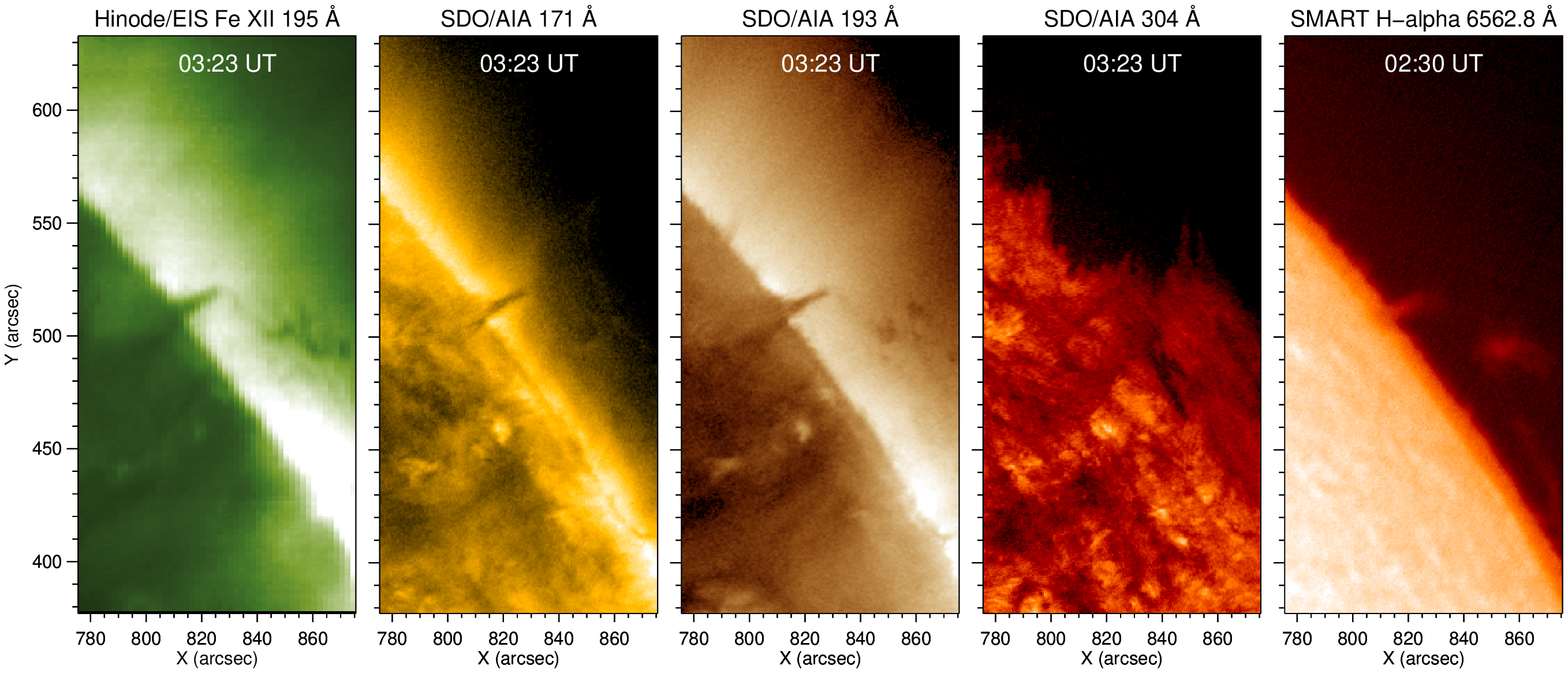}
\caption{The tornado-like prominence observed above the west limb on 2013 September 14th by Hinode/EIS (Fe {\sc xii} 195 {\AA}), SDO/AIA (171 {\AA}, 193 {\AA} and 304 {\AA}) and SMART H-alpha line center.}
\end{center}
\end{figure}

\clearpage

\begin{figure}
\begin{center}
\epsscale{0.8}
\includegraphics[angle=0,scale=.60]{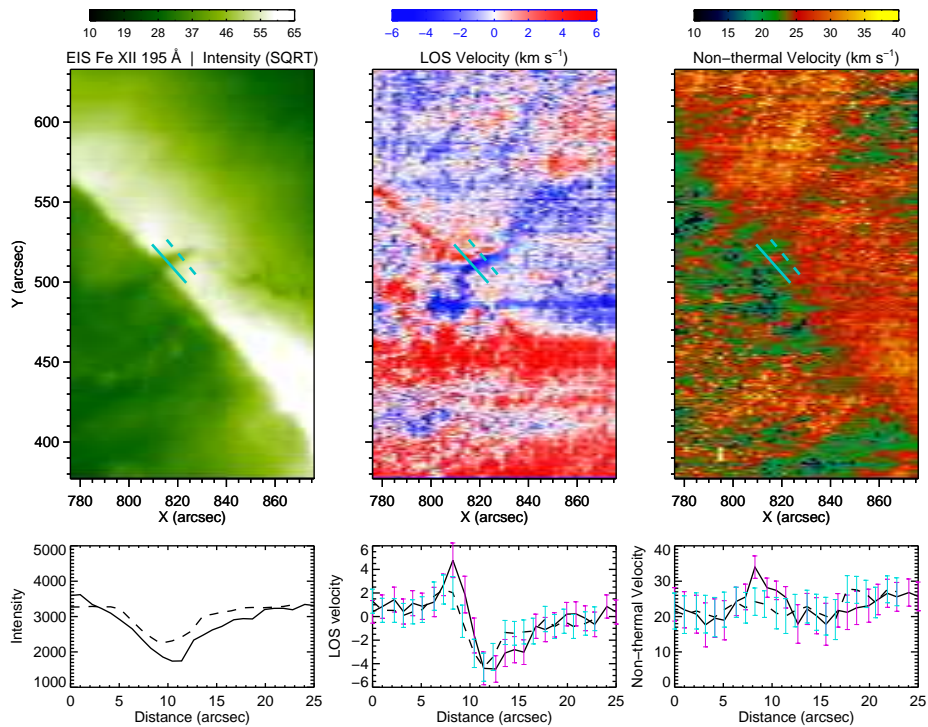}
\caption{Top row: Raster images of EIS Fe {\sc xii} 195 {\AA} intensities, velocities, and non-thermal velocities derived from the spectral fitting. Bottom row: Intensities, LOS velocities, and non-thermal velocities along the two cuts across the barbs as indicated in the top row images (solid and dashed black lines). The error bars on velocities are shown in purple and cyan colors for clarity. \label{fig2}}
\end{center}
\end{figure}

\clearpage

\begin{figure}
\begin{center}
\includegraphics[angle=90,scale=.60]{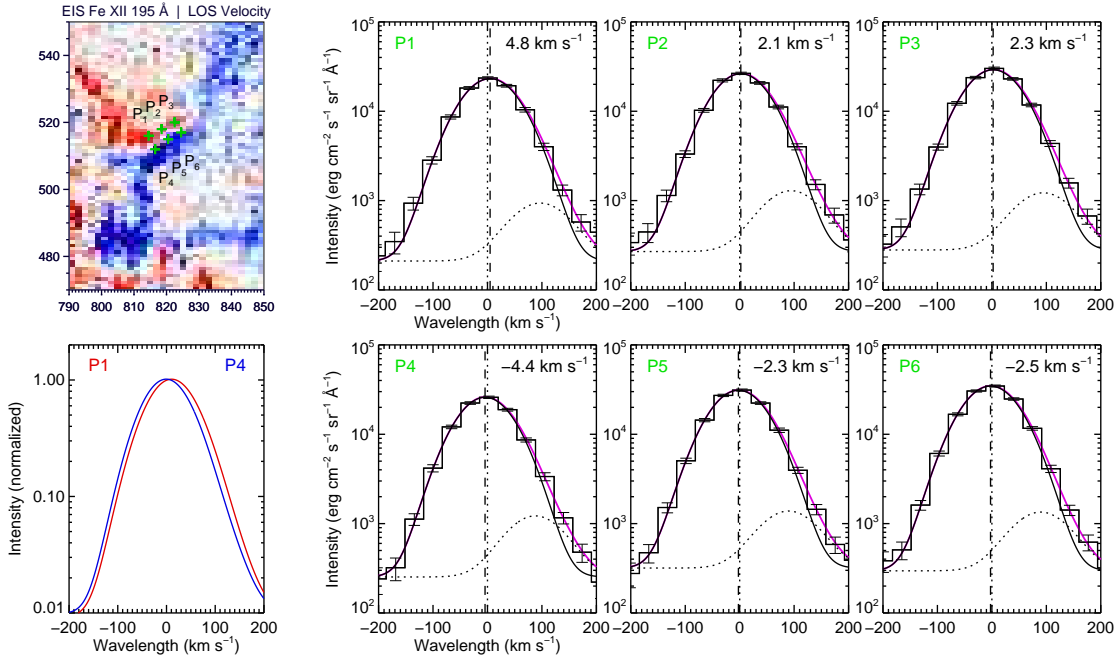}
\caption{EIS Fe {\sc xii} 195 {\AA} line profiles and spectral fitting for selected pixels (P1 to P6). The image on top-left shows the velocity map and six selected pixels. The plot below the image compares the relative shifts between the fitting results of pixels P1 (red) and P4 (blue). The six plots on the right shows detailed spectral fitting for the six pixels by a double-Gaussian function (solid black line and dotted black line). The solid purple line is the total of the two Gaussian functions. Zero velocity is indicated by a dotted vertical line. The estimated Doppler shifts are marked by the dashed line.\label{fig3}}
\end{center}
\end{figure}

\clearpage

\begin{figure}
\begin{center}
\includegraphics[angle= 0,scale=.60]{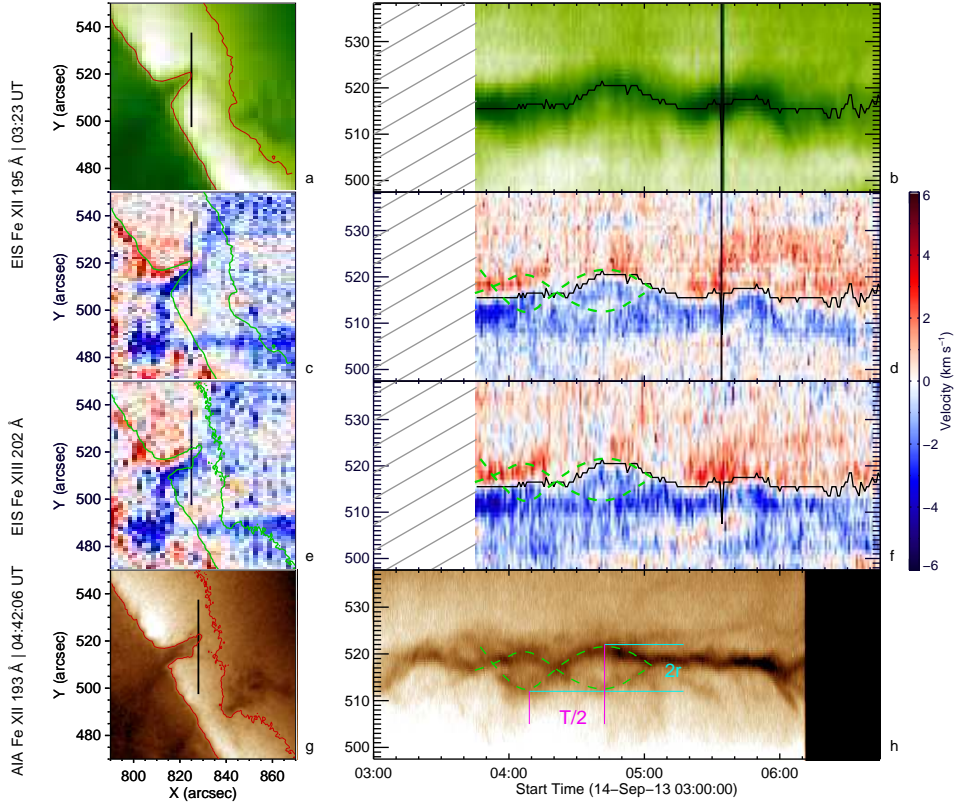}
 \caption{(a) EIS Fe {\sc xii} 195 {\AA} intensity map derived from the scanning mode. The solid black line indicates part of the EIS slit, located at $x =$825${\arcsec}$. (b) Time-distance map of EIS Fe {\sc xii} 195 {\AA} intensity along the EIS slit. The black curve marks the minimum intensity along the slit at each time interval. (c) Velocity map at EIS Fe {\sc xii} 195 {\AA} derived from the scanning mode. (d) Time-distance maps for velocity along the EIS slit. The black curve is the same as in (a). (e) Velocity map at EIS Fe {\sc xiii} 202 {\AA}. (f) Time-distance map of EIS Fe {\sc xiii} 202 {\AA} velocity along the EIS slit. (g) AIA 193 {\AA} map taken at 04:42:06 UT. (h) Time-distance map for AIA 193 {\AA} intensity along the solid black line at $x =$828{\arcsec}. The two green dashed curves show the periodic motions of two thread-like structures. $T$ and $r$ are used to estimate the rotational speed. No effective data was obtained after 06:12 UT (the black block). The contour in (a) and (c) outlines the edge of tornado-like prominence observed by EIS. The contour in (e) and (g) shows the same but for AIA 193 {\AA}. A movie of AIA 193 {\AA} images and the evolution of velocities along EIS slit is available online. \label{fig4}}
\end{center}
\end{figure}

\clearpage

\begin{figure}
	\centering
		\includegraphics[width=0.8\textwidth ]{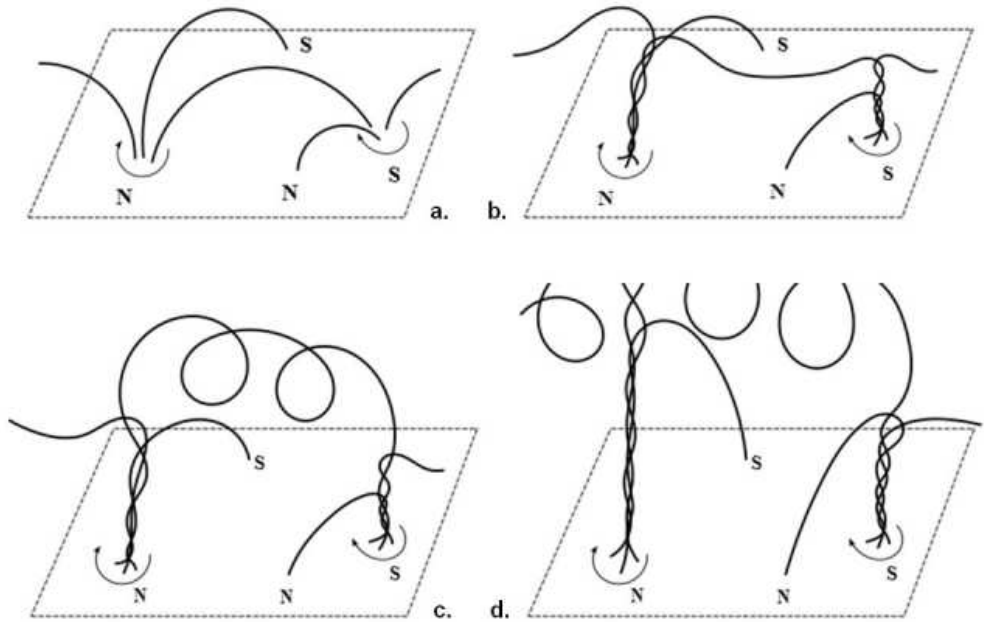}
	\caption{The concept of vortex-filament picture. (a) The filament starts from several vortices located at photospheric networks boundaries. These vortices may be magnetic spots connected with magnetic field lines. (b) Vortex flows concentrate the plasma and magnetic field on the surface and twist the magnetic field lines in filament barbs. (c) Magnetic field lines in the filament spine become twisted to form a flux rope. (d) As vortices continuously twist magnetic field lines, the system may become unstable and erupt. \label{fig:fig5}}
\end{figure}


\end{document}